# SUPERHEAVY ELEMENTS IN THE MAGIC ISLANDS

CHHANDA SAMANTA

*Saha Institute of Nuclear Physics, 1/AF Bidhannagar, Kolkata 700064, India*
*Physics Department, University of Richmond, Richmond, VA 23173, USA*

Recent microscopic calculation based on the density functional theory predicts long-lived superheavy elements in a variety of shapes, including spherical, axial and triaxial configurations. Only when N=184 is approached one expects superheavy nuclei that are spherical in their ground states. Magic islands of extra-stability have been predicted to be around Z=114, 124 or, 126 with N=184, and Z=120, with N=172. However, the question of whether the fission-survived superheavy nuclei with high Z and N would live long enough for detection or, undergo alpha-decay in a very short time remains open. In this talk I shall present results of our calculations of alpha-decay half lives of heavy and superheavy nuclei. Calculations, carried out in a WKB framework using density-dependent M3Y interaction, have been found to reproduce the experimental data quite well. Fission survived Sg nuclei with Z=106, N=162 is predicted to have the highest alpha-decay half life (~3.2 hrs) in the Z=106-108, N=160-164 region called, small 'island/peninsula'. Neutron-rich (N >170) superheavy nuclei with Z >118 are found to have half-lives of the order of microseconds or, less.

## 1. Introduction

In 1965 Myers and Swiatecki [1] showed that the shell corrections added to liquid drop model well reproduces not only the nuclear masses but also the ground state quadrupole moments. More over, it pointed out the possibility of closed shells at Z=114 and N=184 which was later confirmed by A. Sobiczewski, F. A. Gareev, B. N. Kalinkin [2]. In Ref. 1, possibility of other candidates for magic numbers was also stressed. The possible candidates for magic numbers in the Nilsson scheme [3] with the parameters fitted to experimental odd-A nuclei data are Z=114 and N=164 (not N=184). In 1969 Nilsson et al. [4] predicted that the longest fission half-life center rather symmetrically around the nucleon numbers Z=114, N=184. Any stability against spontaneous fission in this region is due to extra binding resulting from the shell effect which essentially increases the alpha half-lives for nuclei with Z<114 and N<184 and decreases those for nuclei with Z>114 and N>184.

In 1969, Mosel and Greiner [5] investigated the dependence of the fission barrier on the level-distributions at the Fermi-surface and calculated lifetimes for superheavy nuclei. Two regions of relative stability against spontaneous fission, which are connected with the so called magic proton numbers $Z=114$ and $Z=164$ were discussed. The nuclei around $Z=164$ were investigated for the first time. The lifetimes for α-decay were also estimated.

In 1972, Fiset and Nix [6] predicted that the nucleus $Z=110$, $N=184$ has the longest total half-life $T \sim 10^{9.4}$ years which is comparable to the age of the earth ($\sim 4.54 \times 10^9$ years). This nucleus was predicted to decay predominantly by α-emission. According to them, as a general rule, the predominant decay mode is α-emission for nuclei containing more than 110 protons, or a few more neutrons than 184; β-emission for nuclei containing less than 110 protons; and spontaneous fission for nuclei containing either less than 184 neutrons or, substantially more. Once the closed proton shell at $Z=114$ is reached, the effect of single particles on the alpha-decay rate is reversed; the alpha decay probability is enhanced for nuclei decaying towards closed shell, while it is hindered for nuclei decaying away. This causes the predominant decay mode to switch to electron capture at $Z=114$, $N=174$. Because of the odd-particle hindrance against alpha-decay and spontaneous fission, and the odd-particle enhancement of electron capture, the nucleus $Z=120$, $N=181$ was predicted to decay predominantly by alpha-decay with α−decay half-life $T_\alpha \sim 5.3$ ms that is greater than the value $T_\alpha \sim 0.48$ ms for the nucleus $Z=120$, $N=182$.

Theoretical predictions of long-lived superheavy nuclei prompted a world-wide search for such nuclei in nature [7], but none has been found so far. In the laboratory however several superheavy nuclei have been produced of which chemically characterized so far are Seaborgium (element 106), Bohrium (element 107), Hassium (element 108) and recently, the element 112 (Ununbium) [8]. Elements 110 and 111 have been named Darmstadtium (in 2003) and Roentgenium (in 2004) respectively. Elements above 111 have not been named as yet.

The shell structure of superheavy nuclei was investigated within various parameterizations of relativistic and non-relativistic nuclear mean-field models [9] and nuclei with ($Z =114$, $N=184$), ($Z=120$, $N=172$) or, ($Z=126$, $N=184$) were found to be doubly-magic. Shell corrections to the nuclear binding energy as a measure of shell effects in superheavy nuclei were studied within the self-consistent Skyrme-Hartree-Fock and relativistic mean-field theories [10]. It was demonstrated that for the vast majority of Skyrme interactions commonly employed in nuclear structure calculations, the strongest shell stabilization appears for *Z*=124 and 126, and for *N*=184. On the other hand, in the relativistic approaches the strongest spherical shell effect appears systematically for *Z*=120 and *N*=172. This difference probably has its roots in the spin-orbit potential. It was also shown that, in contrast to shell corrections which are fairly independent of the force, macroscopic energies extracted from self-consistent calculations strongly depend on the actual force parameterization used. That is, the *A* and *Z* dependence of the mass surface when extrapolating to unknown superheavy nuclei is prone to significant theoretical uncertainties.

Cwiok, Heenen and Nazarewicz [11] considered the interplay between the attractive nuclear force and the disruptive Coulomb repulsion between the protons that favors fission using a self consistent energy density functional theory of Kohn and Sham [12]. They predicted that the long-lived superheavy elements can exist in a variety of shapes, including spherical, axial and triaxial configurations. Only when N=184 is approached one expects superheavy nuclei that are spherical in their ground states. In some cases existence of metastable states and shape isomers can affect decay properties and nuclear half-lives. Later on in a study of $^{254}$No, (Z=102, N=152) by R. -D. Herzberg et al. [11] three excited structures were found, two of which are isomeric (metastable). One of these structures is firmly assigned to a two-proton excitation. These states are highly significant as their location is sensitive to single-particle levels above the gap in shell energies predicted at $Z = 114$, and thus provide a microscopic benchmark for nuclear models of the superheavy elements.

In the beginning of the 1980's the first observations of the elements with Z= 107-109 were made at GSI, Germany [13]. In 1994, α-decay chains were observed from nucleus $^{269}$110 and later on, α-decay chains from nuclides $^{271}$110, $^{272}$111, $^{277}$112, $^{283}$112 were detected at GSI [14]. Recently, the doubly-magic deformed $^{270}$Hs (Z= 108, N = 162) superheavy nucleus has been produced [15]. In Japan, RIKEN reconfirmed the α-decay chains from $^{271}$110, $^{272}$111 and $^{277}$112 and found signature of the nucleus $^{278}$113 [16 -18]. JINR, Russia has reported α−decay chains of several isotopes of nuclei with Z =106-116 and 118 [19-23]. None of these experiments has reached the N=184 region so far.

## 2. Calculations

The α-decay half lives have been calculated extensively through different models [24]. Calculations in the framework of quantum mechanical tunneling of an α-particle from a parent nucleus has been found to provide an excellent description of the experimental data when experimental Q-values are employed along with density-dependent M3Y interaction [25]. Calculation detail of the α-decay half lives of superheavy nuclei in this framework has been described in references [25 - 28]. Only a brief outline of the method is given here.

The required nuclear interaction potentials are calculated by double folding the density distribution functions of the α-particle and the daughter nucleus with density-dependent M3Y effective interaction. The microscopic α-nucleus potential thus obtained, along with the Coulomb interaction potential and the minimum centrifugal barrier required for the spin-parity conservation, form the potential barrier. Total interaction energy at a separation distance R between the daughter and the α-particle is defined as [27],

$$E(R) = V_N(R) + V_C(R) + \hbar^2 \ell(\ell+1) / (2\mu R^2)$$

The WKB action integral is,

$$K = (2/\hbar) \int_{R2}^{R3} [2\mu (E(R) - E_v - Q)]^{1/2} dR \quad \ldots\ldots\ldots (1)$$

At the three "Turning points (TP)",

$$E(R1) = E(R2) = E_v + Q = E(R3).$$

The α-particle oscillates between 1$^{st}$ and 2$^{nd}$ turning points and tunnels through the barrier at 2$^{nd}$ and 3$^{rd}$ TP. The zero point vibration energy $E_v \propto Q$, where $E_v = 0.1045Q$ for even Z-even N, $0.0962Q$ for odd Z-even N, $0.0907Q$ for even Z–odd N, $0.0767Q$ for odd-odd parent nuclei (includes pairing and shell effects) [24]. The decay half life of the parent nucleus is,

$$T = [h \ln 2 / 2E_v] \cdot [1 + \exp(K)] \quad \ldots\ldots\ldots (2)$$

The calculated half lives are very sensitive to Q, as it goes to the exponential function in Eq. (2) through the action integral in Eq. (1). Theoretical Q-values are taken from three different mass formulas: (i) Muntian-Hofmann-Patyk-Sobiczewski (Q-MMM) [29], (ii) Myers-Swiatecki (Q-MS) [30] and, (iii) Koura-Uno-Tachibana-Yamada (Q-KUTY) [31]. Fission half lives are taken from the experimental data and the predictions of Smolanczuk et al. [32]. Beta-decay half lives are taken from Moller-Nix-Kratz [33].

### 3. Results and Discussions

Alpha decay half lives of about 1700 isotopes of elements with $100 \leq Z \leq 130$ have been calculated [26]. Calculations with Q-values from experiment and Q-MMM well reproduce the experimental data for even-even nuclei with $\ell=0$ transition [25-28, 34]. For some odd-odd or, odd A nuclei, $\ell \neq 0$ is needed. At N=184, for Z=110, 112, 114, with Q-KUTY, $T_\alpha \sim 10^{10}$s, $10^8$s, $10^6$s respectively. Theoretical fission half-life values ($T_{SF}$) of these nuclei are $\sim 10^{12}$s, $10^{13}$s, $10^{13}$s respectively. The nuclei $^{296}$112 (N=184), $^{298}$114 (N=184) are β-stable whereas, $^{294}$110 (N=184) is predicted to have large $T_\beta$ in ref. [33] due to its very small positive $Q_\beta$-value.

In the initial RIKEN data for Z=113 (in 2004) [16] there was some discrepancy. As Q value decreases, the half life value should generally increase [27], but an opposite trend was observed in the decay of 111 and 109 (Table 1). There is no discrepancy in the repeat data of RIKEN (in 2007) [16].

Table 1. The α-decay chains of Z=113 (observed in 2004) and $T_\alpha$ calculations.

| Parent Nuclei $^A Z$ | Expt. $E_\alpha$ (MeV) Ref. [16] | Expt. Q (MeV) Ref. [16] | Expt. Decay Time(t) $T_\alpha$=0.693*t | Theory Ref. [27] $T_\alpha$ |
|---|---|---|---|---|
| $^{278}$113 | 11.68 ± 0.04 | 11.90 ± 0.04 | 344 μs (238 μs) | (-18) 10 (+27) μs |
| $^{274}$111 | 11.15 ± 0.07 | 11.36 ± 0.07 | 9.26 ms (6.41 ms) | (-0.12) 0.39 (+0.18) ms |
| $^{270}$109 | 10.03 ± 0.07 | 10.23 ± 0.07 | 7.16 ms (4.96ms) | (-17.68) 52.05 (+27.02) ms |
| $^{266}$107 | 09.08 ± 0.04 | 09.26 ± 0.04 | 2.47 s (1.71 s) | (-1.38) 5.73 (+1.82) s |
| $^{262}$105 | | S.F. | | |

For the $^{277}$112 and its alpha-decay chain, some discrepancies were observed between the GSI and RIKEN data [17]. While the observed first four alpha-decay chains of GSI and RIKEN are similar, the chain-3 of GSI extends up to $^{257}$No. Calculations [28] in a quantum tunneling model, with $\ell=0$, reasonably reproduce the experimental data [17] of $\alpha_2$ and $\alpha_3$ decay channels.

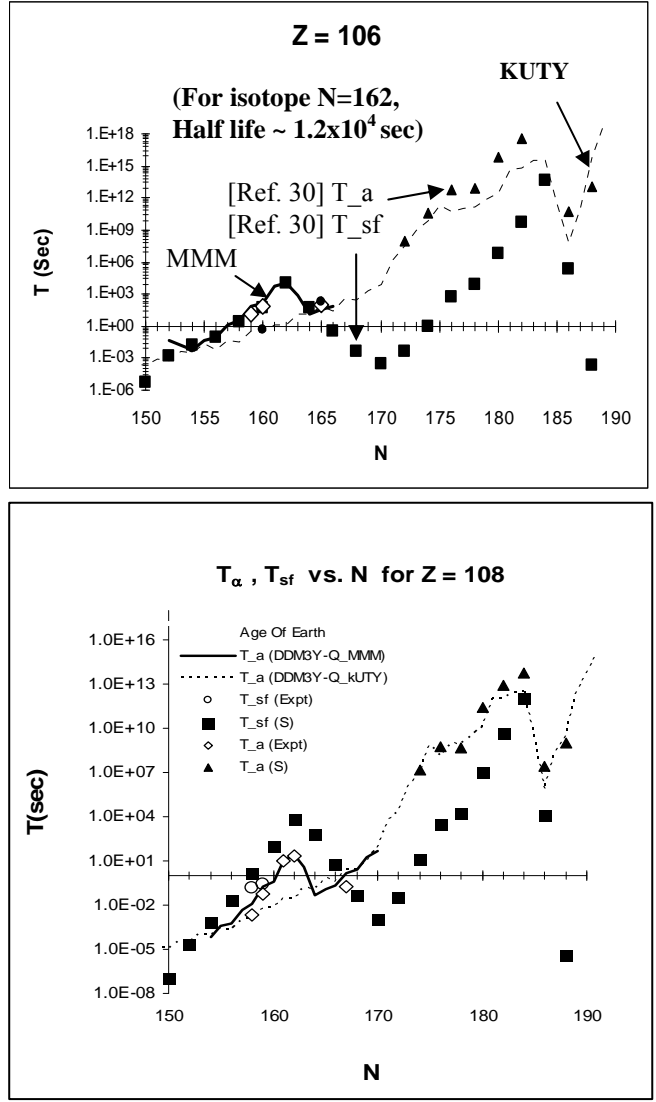

Fig. 1 Plot of α-decay half-life T_a (with Q-values from KUTY [29], dashed lines; MMM [27], solid lines; and Ref. [30], solid triangles) and fission half-life T_sf (filled squares) versus neutron number (N) of Z=106 and 108. Experimental data of half-lives for α-decay [T_a (Expt)] and fission [T_sf (Expt)] are also shown for comparison.

For the $\alpha_1$ decay of $^{277}112$, the experimental data [17] are much higher than the theoretical predictions (with $\ell=0$), but a consideration of higher $\ell$ value can explain the data. On the contrary, for the $\alpha_4$, $\alpha_5$ and $\alpha_6$ decays, the calculated alpha-decay half lives (even with $\ell=0$) are higher [28] than the experimental ones which can not be explained in the present formalism. In this context, further experimental data with higher statistics would be useful.

Detailed investigation [34] indicates that in the region Z =106-108 and N~160-164 a small 'island/peninsula' might survive fission and β-decay, and superheavy nuclei in this region might predominantly undergo α−decay. Interestingly, as shown in Fig.1, in the N~162 region the β-stable, fission survived $^{268}$Sg (Z=106) has $T_\alpha$ ~3.2hrs that is greater than $T_\alpha$ ~28s of the deformed doubly-magic $^{270}$Hs (Z=108). Both $T_\alpha$ are calculated with Q-MMM.

A careful scrutiny [34] also reveals that $^{298}114$ is not the center of the magic island as predicted earlier [4]. On the contrary, the nucleus with Z=110, N=183 appears to be near the center of a possible 'magic island' (Z=104 -116, N~176 - 186) with $T_\alpha$~352 yrs (with Q-KUTY). The nucleus $^{290}$Sg has $T_\alpha$ ~$10^8$ yrs (with Q-KUTY) and $T_{SF}$ ~$10^6$ yrs. Therefore, it might have longer life time compared to other super-heavies. However, for both $^{293}$Ds (N=183) and $^{290}$Sg (N=184) nuclei, β-decay might be another possible decay mode with large $T_\beta$ values.

## 4. Summary

The present scenario of the field of superheavy nuclear research is extremely exciting with the availabilities of new facilities as well as pioneering data in the field. The experimental data are in good agreement with theoretical calculations in a quantum tunneling model with density-dependent M3Y interaction [25-28, 34]. Near N~162, a search for the $^{268}$Sg can be pursued as it might be the longest-lived superheavy nucleus ($T_\alpha$ ~3.2 hrs) in the N=162 region. In the N=184 region, use of Q-KUTY in above formalism indicates some long-lived superheavy nuclei. For example, $^{290}$Sg has $T_\alpha$ ~$10^8$ yrs (although β-decay of this nucleus might be possible). Contrary to earlier prediction [6], the nucleus $^{294}$Ds has $T_\alpha$~311 yrs, a value much less than the age of the earth. For superheavy nuclei with Z >116 and N ~184 the α-decay half-lives are less than one second. In fact, Z=120, 124, 126 with N=184 might form spherical doubly-magic nuclei and survive fission [32] but, they would undergo α-decay within microseconds or, less [26]. While the deformed doubly-magic superheavy nucleus (Z=108, N=162) has been produced in the laboratory, search for the N= 184 spherical doubly-magic nuclei with Z >118 would be a great experimental challenge.

**Acknowledgments**

It is a pleasure to thank my collaborators D.N. Basu and P. Roy Chowdhury. The funding was provided by the Saha Institute of Nuclear Physics, India and the University of Richmond, USA.

**References**

1. W.D. Myers and W.J. Swiatecki, *Report UCRL*, 11980 (1965).
2. A. Sobiczewski, F. A. Gareev, B.N. Kalinkin, *Phys. Lett.* **22**, 500 (1966).
3. S. G. Nilsson, *Mat. Fys. Medd. Dan. Vid. Selsk.* **29**, 16 (1955).
4. S.G.Nilsson, C.F. Tsang, A. Sobiczes, Z. Szymansk, S. Wycech, C. Gustafso, I.L. Lamm, P. Moeller, B. Nilsson *Nucl. Phys.* A**131**, 1 (1969).
5. U. Mosel, W. Greiner, Z. Phys. **222**, 261 (1969); W. Greiner, Int. J. Mod. Phys. E **5**, 1 (1995).
6. E. O. Fiset and J. R. Nix, *Nucl. Phys.* A**193**, 647 (1972).
7. R. K. Bull, *Nature* **282,** 393 (1979); T. Lund, R. Brandt, D. Molzahn, G. Tress, P. Vater and A. Marinov, *Z. für Phys.* A**300**, 285 (1981); K. Murtazaev, V. P. Perelygin, *Sov. At. Energy* **63,** 407 ( 1987).
8. R. Eichler et al., *Nature* **447**, 72 (2007); Andreas Türler, Nature **447**, 47 (2007).
9. K. Rutz, M. Bender, T. Burvenich, T. Schilling, P.-G. Reinhard, J.A. Maruhn and W. Greiner, *Phys. Rev.* C**56**, 238 (1997); W. Greiner, *J. Nucl. Radiochem. Sci* **3**, 159 (2002).
10. A. T. Kruppa, M. Bender, W. Nazarewicz, P.-G. Reinhard, T. Vertse, and S. Ćwiok, *Phys. Rev.* C**61**, 034313 (2000).
11. S. Cwiok, P.-H. Heenen and W. Nazarewicz, *Nature* **433**, 705 (2005); R.-D. Herzberg et al., *Nature* **442**, 896-899 (2006).
12. W. Kohn, L. J. Sham, *Phys. Rev. A***140**, 1133 (1965).
13. P. Armbruster, *Acta Phys. Pol.* B**34**, 1825 (2003); P. Armbruster, *Annu. Rev. Nucl. Part. Sci.* **35**, 135 (1985).
14. S. Hofmann et al., *Euro. Phys. Jour.* A**32**, 251(2007); S. Hofmann, G. Munzenberg, *Rev. Mod. Phys.* **72**, 733 (2000); S. Hofmann, *Rep. Prog. Phys.,* **61**, 639 (1998); G. Munzenberg, *Rep. Prog. Phys.* **51**, 57 (1988).
15. J. Dvorak et al., *Phys. Rev. Lett.* **97**, 242501 (2006).
16. K. Morita et al., *J. Phys. Soc. Jpn.* **73**, 2593 (2004) ; *J. Phys. Soc. Jpn.* **76**, 045001 (2007).
17. K. Morita et al., *J. Phys. Soc. Jpn.* **76**, 043201(2007).
18. K. Morita et al., *J. Phys. Soc. Jpn.* **73**,1738 (2004) ; Euro. Phys. Jour. A**21**, 257 (2004).

19. Yu.Ts. Oganessian, *Euro. Phys. Jour.* D**45**, 17 (2007)**;** *Phys. Rev.* C**76**, 011601(R) (2007); *J. Phys.* G**34**, R165 (2007).
20. Yu.Ts. Oganessian et al., *Phys. Rev.* C**74**, 044602 (2006).
21. Yu.Ts. Oganessian et al., *Phys. Rev.* C**72**, 034611 (2005) ; *Phys. Rev.* C**71**, 029902(E) (2005).
22. Yu.Ts. Oganessian et al., *Phys. Rev.* C**70**, 064609 (2004); *Phys. Rev.* C**69**, 021601(R) (2004).
23. R. Eichler et al., *Nucl. Phys.* A**787**, 373 (2007) ; Yu.A. Lazarev et al., *Phys. Rev.* C**54**, 620 (1996).
24. D.N. Poenaru, W. Greiner, K. Depta, M. Ivascu, D. Mazilu and A. Sandulescu, *Atomic Data and Nuclear Data Tables* **34**, 423-538 (1986); D.N. Poenaru, I-H. Plonski, and Walter Greiner, *Phys. Rev. C***74**, 014312 (2006); D.N. Poenaru, I-H. Plonski, R.A. Gherghescu and Walter Greiner, *J. Phys. G: Nucl. Part. Phys.* **32**, 1223 (2006).
25. P. Roy Chowdhury, C. Samanta and D.N. Basu, *Phys. Rev.* C**73**, 014612 (2006).
26. P. Roy Chowdhury, C. Samanta and D.N. Basu*, Atomic Data and Nuclear Data Tables* (in press), arXiv:0802.4161v2 [nucl-th].
27. C. Samanta, P. Roy Chowdhury and D.N. Basu, *Nucl. Phys.* A**789**, 142 (2007); P. Roy Chowdhury, D.N. Basu and C. Samanta, *Phys. Rev.* C**75**, 047306 (2007).
28. C. Samanta, D.N. Basu and P. Roy Chowdhury, *J. Phys. Soc. Jpn.* **76**, 124201 (2007).
29. I. Muntian, Z. Patyk and A. Sobiczewski, *Acta Phys. Pol.* B**32**, 691 (2001); I. Muntian, S. Hofmann, Z. Patyk and A. Sobiczewski, *Acta Phys. Pol.* B**34**, 2073 (2003); I. Muntian, Z. Patyk and A. Sobiczewski, *Phys. At. Nucl.* **66**, 1015 (2003).
30. W.D. Myers and W.J. Swiatecki, Lawrence Berkeley Laboratory preprint LBL-36803, Dec. (1994); *Nucl. Phys*. A**601**, 141 (1996).
31. H. Koura, T. Tachibana, M. Uno and M. Yamada, KUTY mass formula 2005 revised version, *Prog. In Theor. Phys.* **113**, 305 (2005).
32. R. Smolanczuk, J. Skalski, and A. Sobiczewski, *Phys. Rev.* C**52**, 1871 (1995).R. Smolanczuk *Phys. Rev.* C**56**, 812 (1997).
33. P. Moller, J. R. Nix, and K.-L. Kratz, *Atomic Data Nuclear Data Tables* **66**, 131 (1997).
34. P. Roy Chowdhury, C. Samanta and D.N. Basu, Phys. Rev. C (in press), arXiv: 0802.3837v1 [nucl-th].